\documentclass{article}

\usepackage[final,nonatbib]{NeuRIPS2019/neurips_2019}

\usepackage[utf8]{inputenc} 
\usepackage[T1]{fontenc}    
\usepackage{hyperref}       
\usepackage{url}            
\usepackage{booktabs}       
\usepackage{amsfonts}       
\usepackage{nicefrac}       
\usepackage{microtype}      

\usepackage{amsmath}
\DeclareMathOperator*{\argmax}{arg\,max}

\usepackage{algorithm,algorithmic}
\usepackage{multicol}
\usepackage{changepage}
\usepackage[skins]{tcolorbox}
\usepackage{dsfont}
\usepackage{subcaption}
\usepackage{sidecap}
\usepackage{wrapfig}
\usepackage{xr-hyper}
\usepackage{xcolor}

\def\E{{\rm E}}
\usepackage[title]{appendix}

\allowdisplaybreaks

\makeatletter
\newcommand*{\addFileDependency}[1]{
  \typeout{(#1)}
  \@addtofilelist{#1}
  \IfFileExists{#1}{}{\typeout{No file #1.}}
}
\makeatother
 
\newcommand*{\myexternaldocument}[1]{%
    \externaldocument{#1}%
    \addFileDependency{#1.tex}%
    \addFileDependency{#1.aux}%
}

\myexternaldocument{supplementary}

\title{Emergence of Theory of Mind Collaboration in Multiagent Systems}

\author{%
  Luyao Yuan\\
  \texttt{yuanluyao@ucla.edu} \\
   \And
   Zipeng Fu \\
   \texttt{fu-zipeng@engineering.ucla.edu}\\
   \And
   Linqi Zhou \\
   \texttt{Alexzhou907@gmail.com}\\
   \And
   Kexin Yang \\
   \texttt{ykx1998@ucla.edu} \\
   \And
   Song-Chun Zhu \\
   \texttt{sczhu@cs.ucla.edu}\\
   \AND
    \normalfont{Department of Computer Science}\\
    University of California, Los Angeles \\
}

\begin{document}

\maketitle

\begin{abstract}
Currently, in the study of multiagent systems, the intentions of agents are usually ignored. Nonetheless, as pointed out by Theory of Mind (ToM), people regularly reason about other's mental states, including beliefs, goals, and intentions, to obtain performance advantage in competition, cooperation or coalition. However, due to its intrinsic recursion and intractable modeling of distribution over belief, integrating ToM in multiagent planning and decision making is still a challenge. In this paper, we incorporate ToM in multiagent partially observable Markov decision process (POMDP) and propose an adaptive training algorithm to develop effective collaboration between agents with ToM. We evaluate our algorithms with two games, where our algorithm surpasses all previous decentralized execution algorithms without modeling ToM.
\end{abstract}

\section{Introduction}
Developing effective collaborations in a multiagent system (MAS), when there are more than one learning agents, is a challenging problem. Modeling the joint action for all agents enables single agent reinforcement learning (RL) algorithms to be applied to MAS directly, however, the joint action space grows exponentially as the number of agents increases. Decentralized execution and centralized training algorithms aim to cope with this complexity restriction~\cite{foerster2016learning,sukhbaatar2016learning,foerster2017stabilising,lowe2017multi}, where agents learn policy mapping from local observation history to actions through a centralized critic. Nevertheless, two difficulties need to be addressed in current approaches. First, simultaneous agents updates will cause non-stationary environments and impinge the convergence of these algorithms~\cite{foerster2016riddle,foerster2017stabilising,lowe2017multi,foerster2018bayesian,foerster2018counterfactual}. Second, because a MAS is partially observable to all agents, they must infer each other's mental status to effectively coordinate with each other, either with or without communications. For example, it is important for a group of item collecting robots to infer each other's current target, otherwise, there could be robots aiming for the same item or items collected by no one. 

It has been shown that humans, during an interaction, can reason about others’ beliefs, goals, intentions and predict opponents/partners' behaviors, a capability called Theory of Mind (ToM)~\cite{premack1978does,yoshida2008game,baker2017rational}. In some cases, people can even use ToM recursively, and form beliefs about the way others reason about themselves~\cite{de2015higher}. Thus, to collaborate and communicate with people smoothly, artificial agents must also bear similar potentially recursive mutual reasoning capability. Despite the recent surge of multiagent collaboration modeling~\cite{kinney1998learning,sukhbaatar2016learning,das2017learning,foerster2018counterfactual}, integrating ToM is still a nontrivial challenge. A few approaches attempted to model nested belief of other agents in general multiagent systems, but extensive computation restricts the scale of the solvable problems~\cite{doshi2009monte,han2018learning}. When an agent has an incomplete observation of the environment, it needs to form a belief, a distribution over the actual state of the environment, to take actions~\cite{yoshida2008game,han2018learning}. ToM agents, besides their own beliefs about the state, also model other agents' beliefs, forming belief over beliefs. They can further have beliefs about others' belief over belief, so on and so forth~\cite{doshi2009monte,de2014theory,yoshida2008game,de2015higher,de2017estimating}. The intractability of distribution over distribution makes exact solving for ToM agents' nested beliefs extremely complicated~\cite{doshi2009monte}. Therefore, an approach to acquire the sophistication of high-level recursions without getting entangled into the curse of intractability is needed. 

In this paper, we propose an adaptive training process, following which effective collaboration can emerge between agents with ToM by only modeling one level belief over belief. The complexity of higher level recursions can be preserved by the dynamic evolving of agents' tractable belief estimation functions. Intuitively, for a given agent, we don't simulate its behavior by assuming it has a certain level of recursions, which requires modeling nested beliefs all the way up to the desired level~\cite{doshi2009monte,de2014theory,de2015higher,de2017estimating}. We directly learn a function to approximate its actual belief and how to react accordingly. In cooperative games, this learning becomes mutual adaptation, with controlled exploration rate, improving the performance of the multiagent system~\cite{claus1998dynamics}. Also, in our adaptive training procedure, one agent is trained while the other being fixed, creating a stationary environment for the learning agents. Hence, our algorithm won't be influenced by non-stationarity.

To justify the advantage of ToM agents, we evaluate our algorithm in two-player imperfect information games. In these games, there is a public game state available to all players, and each player has a piece of private information. Each agent maintains a belief about its partner's private state and an estimation of its partner's belief about its private state, namely a belief over belief. The belief over others' private information is obtained with counterfactual reasoning~\cite{fisac2017pragmatic,foerster2018bayesian}. The estimation of the partner's belief about one's private information is learned in centralized training, during which agents share their beliefs to others as supervision. We test our algorithms in two multiagent POMDPs. The first is inspired by the robot-human collaboration in the kitchen setting proposed by Fisac et al.~\cite{fisac2017pragmatic} The second game simulates an appointment scheduling process between two agents, each of whom has private time schedules and wants to choose a commonly available time slot to have a meeting. By regulating the message space of the agents, this game becomes a nontrivial POMDP, requiring mutual reasoning to accomplish. 

We compared our adaptive ToM algorithm with several popular multiagent benchmarks using various flavors of policy gradients~\cite{williams1992simple} and Q-learning~\cite{watkins1989learning}, and achieved robust performance close to SOTA models, which are not only trained in a centralized way, but also have centralized modules like shared Q-functions~\cite{fisac2017pragmatic,foerster2018counterfactual} or meta-agent~\cite{foerster2018bayesian}. Moreover, we found that ToM can facilitate the emergence of more universal protocols decipherable across different groups. Our model beats all benchmarks with a large margin in flexible group assignment experiments. 
\section{Related Work}
Most multiagent RL algorithms have their single agent origins~\cite{hernandez2017survey}. The main challenge of generalizing single agent methods to MAS is the trade-off between complexity and optimality. Centralized execution methods guarantee optimality, but have exponential complexity wrt. the number of agents, while decentralized execution sacrifices performance for simplicity. Thus, the idea of centralized learning but decentralized execution finds its balanced position. Independent Q-based algorithms in MAS, from early works investigating small scale multiagent tabular games~\cite{yang2004multiagent,bu2008comprehensive} to a multiagent generalization of DQN~\cite{foerster2016riddle,tampuu2017multiagent} on large-scale state and action spaces, emphasize decentralized execution. However, the non-stationarity caused by simultaneous agents' updates casts a shadow on the convergence of these algorithms. To relieve the non-stationarity, Foerster et al.~\cite{foerster2017stabilising} proposed stabilizing experience replay by memorizing fingerprints of opponents' policies. Another stream of multiagent algorithms originates from the actor-critic method~\cite{lowe2017multi,foerster2018counterfactual}. They learn a decentralized policy guided by a centralized Q-function. However, the large variance of the policy gradient~\cite{williams1992simple} usually restricts the performance of these algorithms.

On top of the generic Q-learning and policy gradient methods, there are various opponent modeling and communication methods proposed to facilitate coordination and further improve the performance of specific types of tasks. COMA in~\cite{foerster2018counterfactual} aims to solve the credit assignment problem in MAS, but they require agents to have the same reward functions.  \cite{foerster2016learning,sukhbaatar2016learning,mordatch2018emergence} proposed multiagent communication with continuous signals during collaboration. They model communication as another type of actions and have specific modules to control it. Nevertheless, transmitting continuous signals requires channels with large bandwidth, which may not be available when execution. Our model, on the other hand, only needs communication with discrete messages in centralized training to develop effective collaboration in decentralized execution.

Emergence of communication protocol using discrete messages has been explored with various types of communication games~\cite{lazaridou2016multi,havrylov2017emergence,evtimova2017emergent,lazaridou2018emergence}. In these games, agents only communicate with non-suited messages~\cite{wagner2003progress}, namely no agent-environment interactions. Nevertheless, our model integrates agent-agent messages and agent-environment actions to infer other agents' mind and collaborate.

Attempts to integrate ToM in opponent modeling has profound cognitive science origin~\cite{de2014theory,baker2017rational}. LOLA in~\cite{foerster2018learning} learns the best response to evolving opponents. Yet, opponents/partners' real-time believes are not considered into policy. Interactive-POMDP (I-POMDP)~\cite{doshi2009monte,han2018learning} moves one more step forward by actually modeling opponents' mental states at the current moment and integrates other's belief into the agent's policy. However, I-POMDP requires extensive sampling to approximate the nested integration over the belief space, action space and observation space, limiting its scalability. The Bayesian action decoder (BAD)-MDP proposed by Foerster et al.~\cite{foerster2018bayesian} and cooperative inverse reinforcement learning (CIRL) proposed by Fisac et al.~\cite{fisac2017pragmatic} also use counterfactual reasoning in their belief update, but their methods are more centralized in the testing process than ours. The BAD-agent is a super-agent controlling all other agents collectively. Deterministic partial policies can easily reveal agents' private information to the BAD-agent and make it public. CIRL requires that human and robot have their Q-function in common knowledge. Instead, our model doesn't depend on any implicit information flowing between agents during testing and assumes no common knowledge. Wen et al.~\cite{wen2019probabilistic} proposed recursive reasoning policies in MAS to accommodate ToM. The only change in their method from~\cite{lowe2017multi} is to use soft Q-learning~\cite{wei2018multiagent,haarnoja2017reinforcement} and SVGD~\cite{wang2016learning} to estimate other agents' policy conditioning on the estimator's action\footnote{They also assume agents to share the same reward function, as they use the estimator's Q-function to predict other agents policies.}. Our algorithm, which uses explicit belief and belief over belief in our value functions, outperforms their implicit modeling approach in partially observable games.

\section{Background and Setting}
Consider a partial observable multiagent game with $N$ agents. At time $t$ each agent $i\in \{1, ..., N\}$ takes an action from its action space $a^t_i\in A_i$ according to $\pi_i(a|\tau_i^t)$, where $\tau_i^t$ stands for agent $i$'s observation and action history $\{o^0_i, a^0_i, \dots, o_i^t, a_i^t\}$. Here we assume that states of this MAS can be decomposed into physical states and private agent states, namely $\widetilde{S} = S\times\Omega_1\times ...\times\Omega_N$, where $S$ is the environment state space, $\Omega_i$ is the agent $i$'s private state space, eg. agent's goal or intention, and $\widetilde{S}$ is the complete game space. At time $t$ agent $i$'s observation $o_i^t = (O_i(s, a_i^t), \omega_i)$, an observation triggered by its action from the physical state and its private agent state, $\omega_i \in \Omega_i$ which we assume constant until the end of one game. If we denote $\widetilde{s}_{\omega_i}$ as agent $i$'s private state at state $\widetilde{s}$ and $\mathbf{a}^t = \{a_1^t, ..., a_N^t\}$, then we have the state transition function $P(\widetilde{s}^{t+1} | \widetilde{s}^t, \mathbf{a}^t) = T(s^{t+1}|s^t, \mathbf{a}^t)\prod_{i=1}^N\mathbf{1}(\widetilde{s}^{t+1}_{\omega_i} = \widetilde{s}^t_{\omega_i})$. At time $t$, agent $i$ gets reward $r_i^t = r_i(\widetilde{s}^t, \mathbf{a}^t)$. In fully cooperative games, all agents share the same reward function. Notice that rewards depend on the complete state instead of just the environment. An example will be some of the agents know the goal of a task requiring all agents' effort to accomplish, making inference of other agents' private state crucial.

We can define the value of a state as $V^{\mathbf{\pi}}(\widetilde{s}^t) = \E_{\mathbf{a}^t, \widetilde{s}^{t+1}, ...}[\sum_t\gamma^tr(\widetilde{s}^t, \mathbf{a}^t)]$. The goal of this multiagent game is to find a set of policies $\mathbf{\pi} = \{\pi_1, ...,\pi_N\}$ to maximize the expected return of the game $\E_{\widetilde{s}^0}[V^{\mathbf{\pi}}(\widetilde{s}^0)]$ starting at time 0. The optimal value of a state is defined as $V^*(\widetilde{s}) = \max_{\pi}V^{\pi}(\widetilde{s})$. We can also define Q-function as $Q^{\pi}(\widetilde{s}^t, \mathbf{a}) = r(\widetilde{s}^t, \mathbf{a}) + \gamma\E_{\widetilde{s}^{t+1}\sim P(\widetilde{s}^{t+1}|\widetilde{s}^t, \mathbf{a})}[V^{\pi}(\widetilde{s}^{t+1})]$ and $Q^{*}(\widetilde{s}^t, \mathbf{a}) = r(\widetilde{s}^t, \mathbf{a}) + \gamma\E_{\widetilde{s}^{t+1}\sim P(\widetilde{s}^{t+1}|\widetilde{s}^t, \mathbf{a})}[V^{*}(\widetilde{s}^{t+1})]$. It has been shown that Q-learning can converge to optimal Q-values with mild assumptions~\cite{jaakkola1994convergence}. We also define value functions for an agent given its opponents policies:
\begin{align}
    \label{eq:v-value}
    V_{i|\pi_{-i}}^*(\widetilde{s}^t)=&\max_{\pi_i}\E_{\mathbf{a}^t_{-i}\sim\pi_{-i}, a_i^t\sim\pi_i, \widetilde{s}^{t+1},...}\Big[\sum_t\gamma^tr_i(\widetilde{s}^t,\mathbf{a}^t)\Big]\\
    \label{eq:q-value}
    Q^{*}_{i|\pi_{-i}}(\widetilde{s}^t, a_i) &= \underset{\mathbf{a}_{-i}\sim\pi_{-i}}{\E}\Big[r_i(\widetilde{s}^t, \mathbf{a}_{-i}, a_i) + \gamma\underset{P(\widetilde{s}^{t+1}|\widetilde{s}^t,\mathbf{a})}{\E}[V^{*}_{i|\pi_{-i}}(\widetilde{s}^{t+1})]\Big].
\end{align}
When there is no ambiguity, we can omit the $\pi_{-i}$ in the subscript and only write $Q_i$.

\section{Method}
\begin{SCfigure}
    \centering
    \caption{\small{Execution pipeline for the case of two agents. Actions are determined by Q-values weighted by the belief. Observation of other agents' actions is utilized to update one's belief by counterfactual reasoning. Dashed lines represent belief estimation supervision, only included in centralized training. Remember that this supervision is a discretization of the real belief, eg. a sample from the belief distribution.}}
    \includegraphics[width=0.5\textwidth]{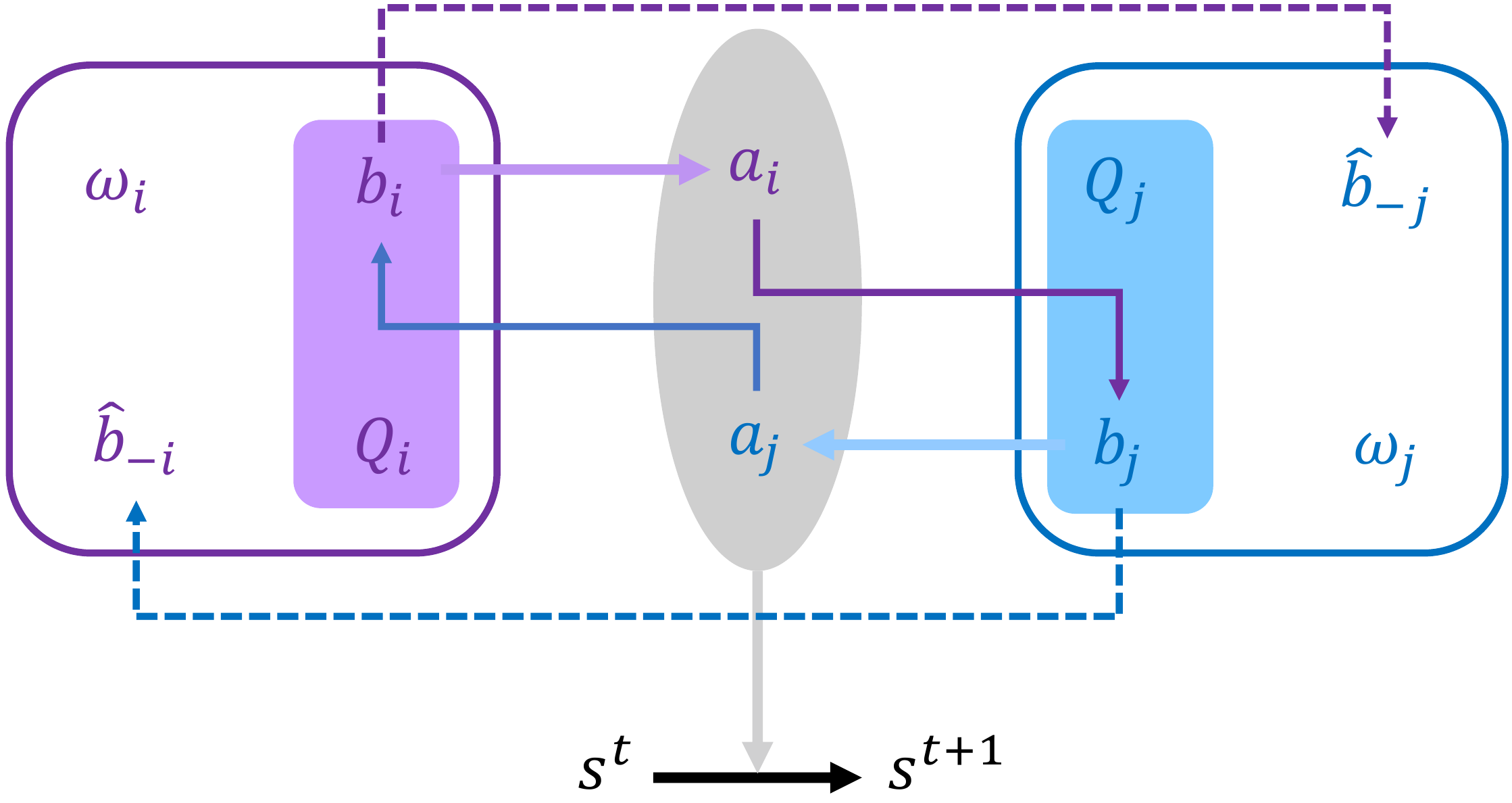}
    \label{fig:pipeline}
    \vspace{-10pt}
\end{SCfigure}
In this section, we propose our algorithm to solve the partial observable game defined in the previous section. First, we define agents' modules and introduce how they act when all modules are learned. Then, we illustrate the learning algorithm for all modules, which is a centralized process among agents.
\subsection{Decentralized Execution}\label{sec:dectralized_exec}
The execution process is purely decentralized as all agents only act according to their local observations and no direct communication other than mutual inference from action observation. For an agent $i$, it performs according to its value function, $Q_i: \widetilde{S} \times A_i \rightarrow R$. Here we assume the physical state is fully observable to all agents, so the only belief an agent needs to hold is about other agents' private states. This assumption can be relieved by introducing an observation model for the physical state and form belief over the environment. If we denote other agents as $-i$, as their private states are not available to agent $i$, it must maintain a belief $b_i^t(\omega_{-i})$. Thus, we have agent $i$'s policy following Boltzmann rationality model~\cite{baker2017rational} with $\beta$ as the rationality coefficient of $i$ and quantifies the concentration of $i$'s choices around the optimum. As $\beta \rightarrow \infty$, $i$ becomes a perfectly rational agent, while, as $\beta \rightarrow 0$, $i$ becomes indifferent to Q:
\begin{align}
    \label{eq:softmax}
    &\pi_i(a|s^t, \omega^t_i, \tau_i^t) = \dfrac{\exp(\beta\sum_{\omega_{-i}\in\Omega_{-i}}b_i^t(\omega_{-i}|\tau_i^t)Q_i(a, s^t, \omega_i, \omega_{-i}))}{\sum_{a'\in A_i}\exp(\beta\sum_{\omega_{-i}\in\Omega_{-i}}b_i^t(\omega_{-i}|\tau_i^t)Q_i(a', s^t, \omega_i, \omega_{-i}))}.
\end{align}
Next problem is how agent $i$ maintains its belief given its observation history. We utilize counterfactual reasoning~\cite{fisac2017pragmatic,foerster2018bayesian} in our belief update function. That is, agent $i$ traverses all possible private states $\omega_{-i}$ and estimates how likely the actions it observed are taken given a specific set of private agent states are the correct one. Then, $i$ updates its belief using Bayesian rule:
\begin{align}
    \label{eq:belief_update}
    b^t_i(\omega_{-i}|\tau_i^t) &= P(\omega_{-i}|\tau_i^{t-1}, \mathbf{a}^t_{-i})\propto \hat{\pi}_{-i}(\mathbf{a}^t_{-i}|s^t, \omega_{-i})b^{t-1}_i(\omega_{-i}),
\end{align}
where $\hat{\pi}_{-i}$ is agent $i$'s estimation of $-i$'s policy, learned in centralized training. In this paper, we assume all actions are observable to all agents. That is, the only hidden components of the games are the agents' private state. This assumption can be relaxed by including an observation function for each agent in equation \ref{eq:belief_update}.

Agent $i$ needs to maintain a belief about other agents' private states, so do other agents need to estimate $i$'s state. If there are two proper actions for task completion, but one can convey agent $i$'s private state to others while the other reveals little information, then the first action should be preferred. Thus, agent $i$'s Q-function should have another argument to accommodate others' belief about $\omega_i$. Utilizing the obverter technique~\cite{choi2018compositional}, we let the $i$ holds a belief $\hat{b}_{-i}$ as the estimation of $-i$'s belief about $\omega_{i}$. Here, $\hat{b}_{-i}$ is still a distribution over $\Omega_i$. We didn't use a distribution over distribution to model this nested belief because the belief update process is deterministic for rational agents following Bayesian rule. Given $\hat{b}_{-i}^0$ a uniform distribution over candidates, $P(\hat{b}_{-i}^t)$ is unimodal with uncertainty merely from the likelihood and can be approximated with a single point. All we need is a belief update function in $i$, $f_{-i}: \Delta(\Omega_i) \times A_i \times S \rightarrow \Delta(\Omega_i)$, where $\Delta(\Omega_i)$ represents a distribution over $\Omega_i$. Belief update function $f$ takes in the old belief, agent $i$'s action, the physical state and returns a new belief as $i$'s new estimation of others' belief over $\omega_i$. If $\omega_i$ and $\omega_{-i}$ are not independent, then $b_i$ should be an additional argument of $f$, but in this paper, we only explore the scenarios where $\omega_i$ and $\omega_{-i}$ are independent of each other. The ability to correctly infer others' private states from their actions and predict others belief about oneself introduces ToM into our agents. Figure \ref{fig:pipeline} visualizes the execution pipeline. 

\subsection{Centralized Training}
\begin{algorithm}[t!]
\footnotesize
\caption{Adaptive ToM Collaboration Emergence}
\label{alg1}
\small{
\begin{algorithmic}[1]
   \small
    \STATE Randomly initialize $\theta_i, \hat{\theta}_i, \eta_Q, \eta_{\pi}, \eta_f, i
    \in \{1, ..., N\}$\\
    \STATE Learning rate $\eta$, Batch size $M$
    \FOR{each round}
        \FOR{$i \in \{1, ..., N\}$}
            \STATE Initialize replay buffer $\mathcal{D} \leftarrow \emptyset$
            \WHILE{train agent $i$}
                \REPEAT
                    \STATE Agents sample actions according to equation \ref{eq:softmax}
                    \STATE Agents update their beliefs
                    \STATE Agents update their estimation of partners' beliefs
                \UNTIL{game ends}
                \STATE Update $\mathcal{D}$ with new trajectory
                \STATE Sample $M$ trajectories $\{(\omega_{1:N}, s_{0:T}, \mathbf{a}_{0:T}, r_{0:T})\}_{k=1}^M$
                \STATE $y_i^{t, (k)} = r_i^{t, (k)} + \gamma \max\limits_{a\in A_i}Q_{\hat{\theta}_i}(a, s^{t+1, (k)}, \mathbf{\omega}^{(k)}, \hat{b}_{-i}^{t, (k)})$
                \STATE $L^Q = \sum\limits_{t, k} ||Q_{\theta_i}(a_i^{t, (k)}, s^{t, (k)}, \mathbf{\omega}^{(k)}, \hat{b}_{-i}^{t, (k)}) - y_i^{t, (k)}||^2$\label{alg:L_q}
                \STATE $L^{\pi} = \sum_{t, k} H(\hat{\pi}_{-i}(\mathbf{a}_{-i}^t|s^{t, (k)}, \omega_{-i}^{t, (k)}), \mathbf{a}_{-i}^{t, (k)})$\label{alg:L_pi}
                \STATE $L^f = KL(\bar{b}_{-i}^{t, (k)}||f_{-i}(\hat{b}_{-i}^{t-1, (k)}, a^{t, (k)}_i, s^{t, (k)}))$\label{alg:L_f}
                \STATE $\theta_i \leftarrow \theta_i - \nabla_{\theta_i}(\eta_Q L^{Q} + \eta_{\pi} L^{\pi} + \eta_f L^f)$\label{alg:update}
                \STATE Periodically update $\hat{\theta}_i \leftarrow \theta_i$ for Q-learning
            \ENDWHILE
        \ENDFOR
    \ENDFOR
\end{algorithmic}
}
\end{algorithm}
As introduced in section \ref{sec:dectralized_exec}, there are three components that every agent needs to learn during centralized training, $Q_i, \hat{\pi}_{-i}$ and $f_{-i}$. To learn $Q_i$, we apply the deep Q-learning algorithm~\cite{watkins1989learning,mnih2013playing}. To avoid the non-stationarity caused by simultaneous agent updates, we fix all other agents when one agent is being trained. Thus, all the fixed agents can be considered as part of the environment, so the convergence of the Q-function of the learning agent is still guaranteed. In a fully observable multiagent game where all agents share the same reward, suppose we denote $\pi = \{\pi_1, ..., \pi_{i-1}, \pi_{i}, \pi_{i+1}, ..., \pi_N\}$ as the joint policy before agent $i$ is being trained and $\pi' = \{\pi_1, ..., \pi_{i-1}, \pi'_{i}, \pi_{i+1}, ..., \pi_N\}$ as the new policy after $i$ is trained, where $\pi_i' = \argmax_{a\in A_i}Q_i^*(s, a)$. One can show from the convergence of the Q-learning that $E_{s_0}[V^{\pi'}(s_0)] \geq E_{s_0}[V^{\pi}(s_0)]$ using the same procedure as in~\cite{jaakkola1994convergence}. Therefore, our adaptive training algorithm guarantees monotonic performance improvement. As history is important in Q-functions of POMDP, we saved trajectories in the replay buffer and sample complete trajectories in the training~\cite{hausknecht2015deep}.

Both $\hat{\pi}_{-i}$ and $f_{-i}$ are learned with supervision. To be more specific, during training agents $-i$ will reveal their private agent states to the public after each game is finished. Then $\hat{\pi}_{-i}$ can be learned with MLE. The supervision of $f_{-i}$ is more complicated, as beliefs are continuous variables representing distributions, but communicating discrete messages are more realistic. Therefore, we discretize beliefs $b_{-i}$ and pass that discretization $\bar{b}_{-i}$ to $i$ as supervision for $f_{-i}$. The exact discretizing procedure may vary according to different tasks, as we'll elaborate in later section \ref{sec:kitchen} and \ref{sec:calendar}. In our future work, we consider using a specifically trained message decoder to further enrich communication. The detailed learning process is in algorithm \ref{alg1}. Here we abuse the notation a little by encompassing all parameters of agent $i$ into $\theta_i$, which contains three sets of parameters for the Q-function, $\hat{\pi}_{-i}$ and $f_{-i}$ respectively. In line \ref{alg:L_pi}, $H$ represents softmax cross-entropy loss, as here we have $A_{-i}$ as discrete actions. In line \ref{alg:L_f} we use Kullback-Leibler (KL) divergence to measure the distance between the predicted partners' belief and the partners' belief supervision, $\bar{b}_{-i}$. This distance measure can be replaced with other functions given different tasks and forms of belief supervision because we use discretized belief instead of continuous vectors. As in line \ref{alg:update}, different loss may use different step size $\eta$ to update.
\section{Experiments}
\begin{figure*}[t]
    \centering
    \begin{subfigure}[b]{0.45\textwidth}
        \includegraphics[width=\textwidth]{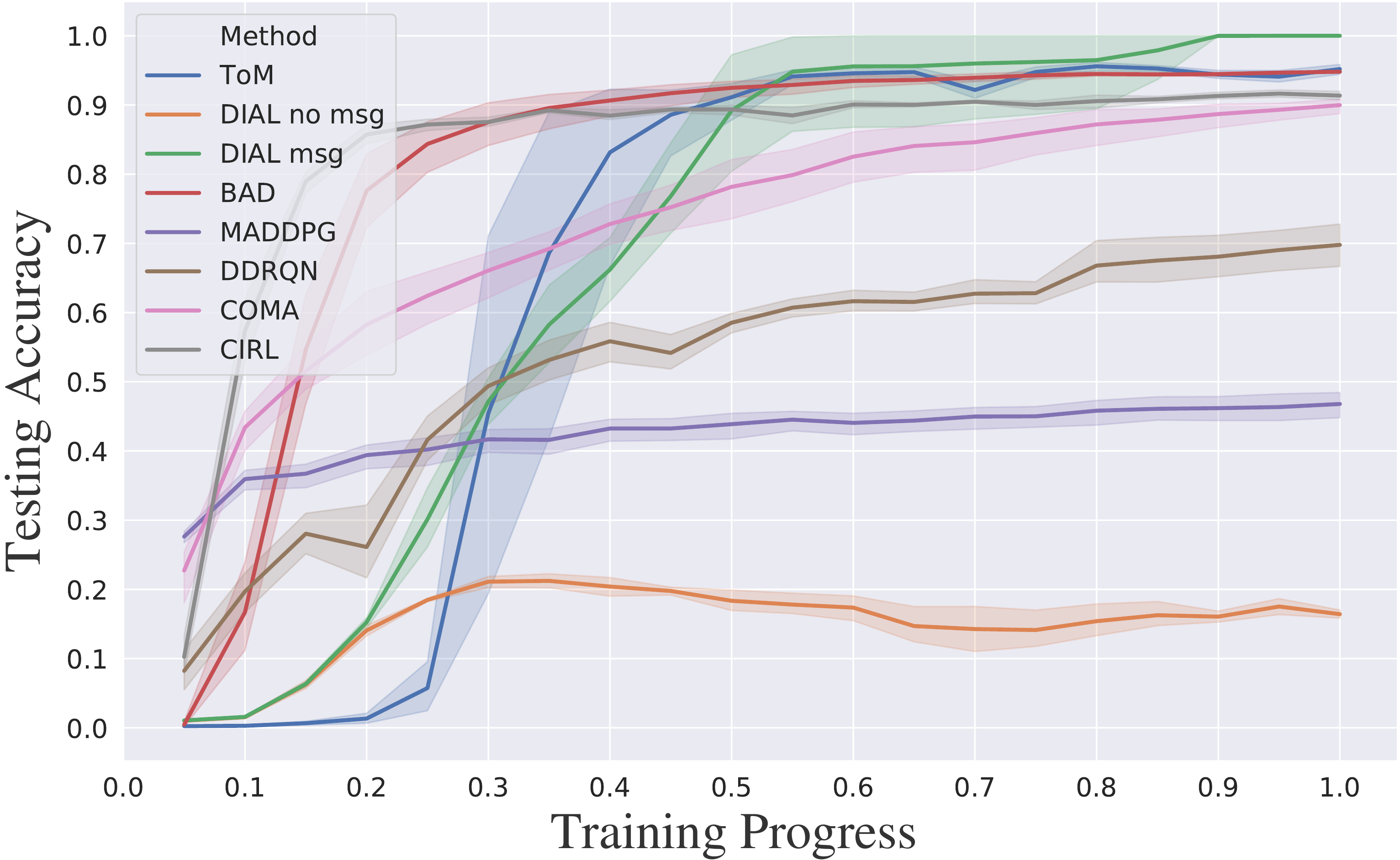}
        \caption{\small{Kitchen Collaboration. BAD:0.15M, COMA:0.25M, ToM: 1.6M, DIAL:0.1M, MADDPG:3M, DDRQN:4M, CIRL:0.4M. When the number of possible states is smaller than the number of actions, BAD can easily form an one-to-one mapping to force agents to reveal their private information. When there are fewer actions than possible states (e.g. in appointment scheduling), BAD has a worse performance.}}
        \label{fig:kitchen4}
        \vspace{-5pt}
    \end{subfigure}~~
    \begin{subfigure}[b]{0.45\textwidth}
        \includegraphics[width=\textwidth]{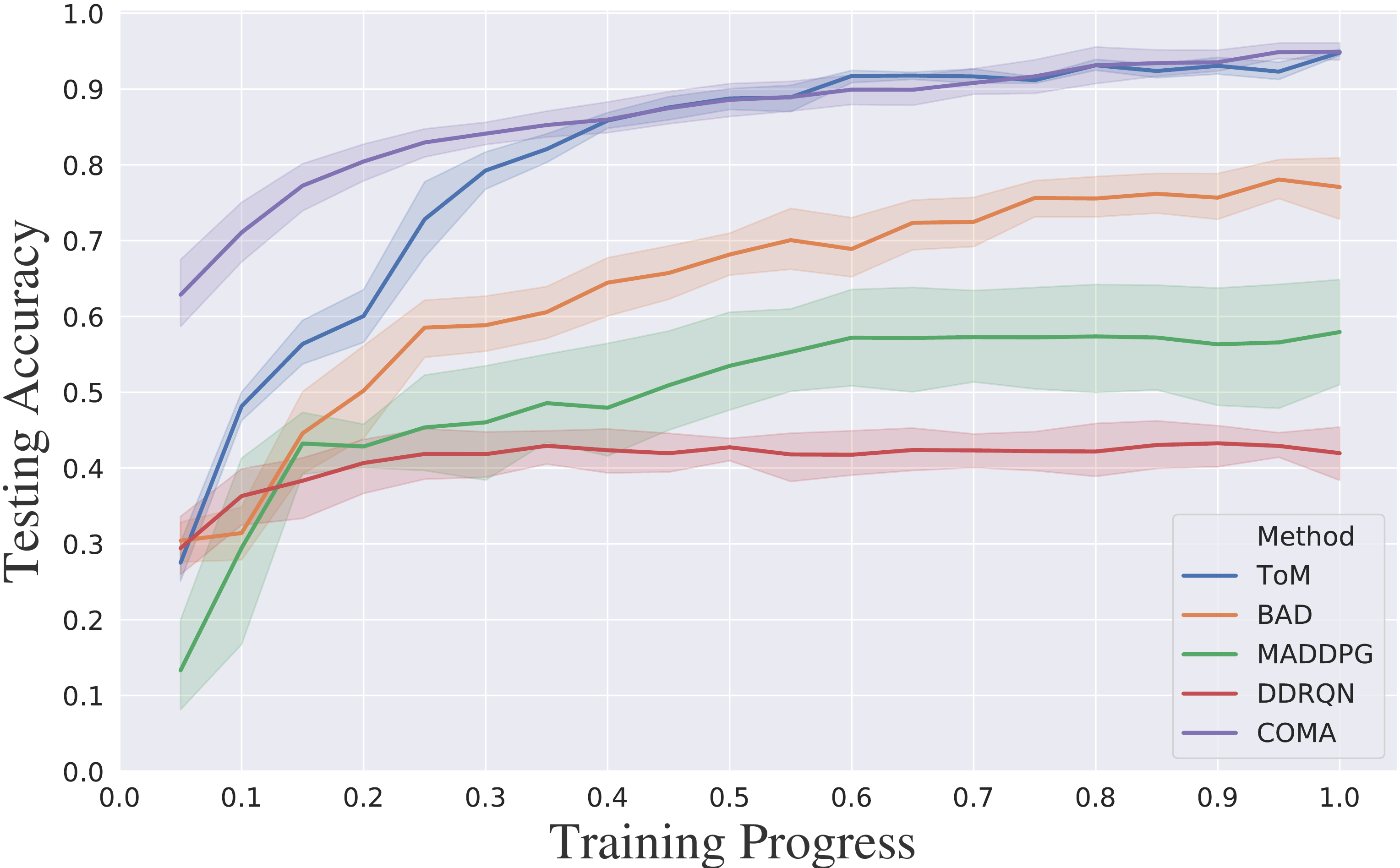}
        \caption{\small{Appointment Scheduling. BAD: 0.1M, COMA: 0.1M, ToM: 1.6M, MADDPG: 1.0M, DDRQN: 20K. Using two separate Q-functions causing MADPPG more likely to have at least one agent stuck at local optimum. DDRQN updates all agents simultaneously, suffering from non-stationarity. Our model cannot outperform COMA probably due to the difficulty of learning high dimensional belief precisely.}}
        \label{fig:calendar8}
        \vspace{-5pt}
    \end{subfigure}
    \caption{\small{Different algorithms take different training iterations to converge, so we normalize the iterations as training progress from 0 to 1 as the x-axis. y-axis represents testing accuracy. Due to the time limit, we stop training a model after it converges. We halt when the best training accuracy stops increasing for 10000 iterations. We average the training accuracy of mini-batches every 1000 iterations to avoid fluctuation. We'll report results for training all models to a fixed iterations in our later version. In captions above, M stands for $10^6$ and K stands for $10^3$.}}
    \label{fig:curves}
    \vspace{-15pt}
\end{figure*}
We used two multiagent games with imperfect information including two agents to evaluate our experiments. For both experiments, we generated three exclusive training/testing splits and repeat two experiments with different random initialization on each split. We report average performance from six experiments. We compare with several benchmarks including CIRL~\cite{fisac2017pragmatic}, COMA~\cite{foerster2018counterfactual}, DDRQN~\cite{foerster2016riddle}, MADDPG~\cite{lowe2017multi}, BAD~\cite{foerster2018bayesian} and DIAL~\cite{foerster2016learning}. For MADDPG, we integrate PR2's policy modeling from~\cite{wen2019probabilistic} by taking history into the policy network. We further simplify policy estimation by letting agents using their opponents' actual policies in training. Notice that in the original setting of DIAL, agents require communication in both training and testing. Our model, on the other hand, only needs communication during training. So we report DIAL performance for testing without communication. When implementing these benchmarks, we stick to their papers to adapt all models to our tasks. We also take the author released code for reference if there is. The detailed structure and hyper-parameters can be found in supplementary together with our codes.
\subsection{Kitchen Collaboration}\label{sec:kitchen}
This multiagent game is inspired by the ChefWorld proposed in~\cite{fisac2017pragmatic}. Suppose there are two agents $A$ and $B$, with $A$ as the chef and $B$ as the assistant. For each game, a recipe with $K$ dishes is publicly provided to the agents. The purpose of the game is for both agents to prepare ingredients for the target dish, which is only known by the chef. We use a categorical number to denote an ingredient and represent every dish on the recipe as a set of ingredients. Agents take turns to act by selecting an ingredient and put it on the workplace, starting from the chef, $A$. A game ends if either a wrong ingredient is selected or all ingredients are prepared. We allow repetitive ingredients in a dish but preparing more than enough also leads to failure. Suppose we have a recipe, $[0, 1, 2, 6]$, $[2, 2, 4, 8, 9]$, $\textbf{[2, 3, 6, 7, 7]}$ and $[1, 2, 2, 8, 9]$, and the target dish is the third dish. Then a successful action sequence can be $[3_A$, $7_B$, $6_A$, $2_B$, $7_A]$\footnote{This is a trajectory generated by our model, notice the unique identifier of the target dish is selected by the teacher in the first round. Agents' usage of unique identifier is shown in table \ref{tab:UI-usage}.}. The order of the ingredients being prepared and who prepared what ingredient does not matter for the completion of the task. Some failure trajectories can be $[2_A$, $0_B]$ or $[3_A$, $7_B$, $7_A$, $7_B]$. At the end of each game, both agents get a reward $\pm1$ for success or failure, and there is no step reward or cost as the game proceeds. In this experiment, agent $B$'s beliefs are $k$-dimensional distribution vectors. Hence, in centralized training, we let agent $B$ sample an index $k$ between $1$ to $K$ from its belief and send a one-hot vector with the $k$-th number being 1 to $A$ as a discretized belief supervision. KL-divergence is used as the distance function for $L^f$. 

There are $W$ ingredients in total and we limit the maximum number of ingredients in a dish by $M$. In a recipe, all dishes are different from other dishes by at least one ingredient. A target dish is a one-hot $K$-dimensional vector, indicating which dish should be prepared. Only the chef knows the target, so we have $\omega_A$ as a $K$-dimensional one-hot vector and $\omega_B$ as an empty dumpy variable. In later of the paper, we refer a game as the combination of a recipe and a target. 

All recipes are randomly generated. We synthesize a dish by randomly picking $M$ numbers from $\{0, 1, ..., W\}$ with replacement. If $W$ is selected, we remove it from the dish as only $\{0, 1, ..., W-1\}$ are valid ingredients. We include $W$ in the selection process to have dishes with fewer than $M$ ingredients. The order of the ingredients doesn't matter, and we will resample a dish if two dishes are duplicates in one recipe. There are 3335 possible dishes and $5.1 \times 10^{12}$ possible games in total. To verify the robustness of our model, we keep the training dishes exclusive from the testing dishes. There are 2335 unique dishes in the training set and 1000 in the testing set. There are 7 and 3 million games in our training and testing set respectively. We evaluated our model and benchmarks using $K = 4, M = 5, W = 10$.

Results are shown in table \ref{tab:results} and figure \ref{fig:kitchen4}. BAD~\cite{foerster2018bayesian} achieves very close performance with our model. Their model has a BAD-agent that controls all other agents by sending partial policies to agents and using equation 1 in their paper to infer agents' private observations. Nevertheless, when the private observation space is small, the BAD-agent can form an one-to-one mapping between actions to private observations, forcing agents to reveal their private states by doing actions as told. Then the private states can flow among all agents through the BAD-agent, centralizing the execution process. In other words, their agents don't do counterfactual inference but tell their private states to the BAD-agent, who determine the joint action as a meta-agent. We refer it as a weakly centralized execution. One drawback of their work is that mappings between actions and private states are usually arbitrary, so the emerged protocols are exclusive to one group of agents. To test our hypothesis, in section \ref{sec:switch}, we conduct switch group experiments. The significant performance reduction illustrates the sub-optimality of the protocols developed by Foerster el al.~\cite{foerster2018bayesian}

In table \ref{tab:UI-usage}, we also report the probability of the chef using the unique ingredients in the target dish in the first round. Using the unique identifier to disambiguate distractors is very common in human communication~\cite{de2015higher,de2017estimating}. Our model developed a protocol very compatible with human communication. In the future, we consider adding human-agent experiments to quantify the compatibility. Using unique ingredients to identify the target dish is not the only possible protocol between collaborative agents, but it is the most universal way across different groups.
\begin{table}
    \centering
    \begin{tabular}{c|c}
        \toprule
        BAD & 39.56 $\pm$ 2.62 \\
        CIRL & 38.20 $\pm$ 0.41\\
        COMA & 33.26 $\pm$ 1.35\\
        DDRQN & 34.19 $\pm$ 1.32 \\
        MADDPG & 34.49 $\pm$ 0.26\\
        ToM-noBoB & 53.52 $\pm$ 2.60\\
        ToM & \textbf{62.46 $\pm$ 3.79}\\
        \bottomrule
    \end{tabular}
    \caption{\small{Unique identifier using percentage. ToM agents can simulate partners' reaction of self actions and choose the most ideal action to perform. In kitchen collaboration, the chef needs to clarify the target dish to the assistant as soon as possible so that only the correct ingredients are prepared. The most straightforward way is for the chef to choose the unique ingredient only included by the target dish. We calculate the probability for the chef to choose a unique ingredient if there is one. We claim that the frequency to choose the unique ingredient is related to the switch group accuracy in \ref{sec:switch}. As if the chef doesn't use unique ingredients to indicate the target dish, there have to be some other protocols formed between agents, which are usually group-specific.}}
    \label{tab:UI-usage}
    \vspace{-15pt}
\end{table}
\begin{table}[ht!]
    \centering
    \caption{\small{Accuracies for kitchen collaboration and appointment scheduling tasks. In the first experiment, BAD has slightly better performance, but it is weakly centralized execution as the private agent states can flow among agents through the BAD-agent. COMA achieves 0.1\% higher accuracy than ours in the second experiment. There is a centralized critic in COMA, so agents must have the same reward function. In section \ref{sec:switch} we show that both of these methods learn group-specific protocols, while ToM can emerge task level protocol. DIAL with messages can achieve perfect performance as agents will share private states directly through the communication channel. We include this algorithm as an oracle. If we remove the communication channel from DIAL, its performance degrades considerably. As the communication channel enables agents to collide easily in appointment scheduling, DIAL is not used for the second experiment. CIRL is applicable when only one agent has hidden information. To show the convergence of the models, test curves are plotted in figure \ref{fig:kitchen4} and \ref{fig:calendar8}.}}
    \begin{tabular}{c|c|c}
        \toprule
         & Kitchen & Appointment \\
        \midrule
        BAD & $\mathbf{95.47 \pm 0.88}$& 78.27 $\pm$ 3.57\\
        CIRL & 91.36 $\pm$ 0.80 & N/A\\
        COMA & 89.98 $\pm$ 1.58 & $\mathbf{94.90 \pm 1.65}$\\
        DDRQN & 69.77  $\pm$ 4.30 & 41.97 $\pm$ 4.91\\
        DIAL & 100.0 $\pm$ 0.00 & N/A\\
        DIRL no msg & 17.99 $\pm$ 1.17 & N/A\\
        MADDPG & 46.79 $\pm$ 2.45 & 57.94 $\pm$ 9.27\\
        Random & 5.67 $\pm$ 0.27 & 25.00 \\
        ToM & $\mathbf{95.19 \pm 1.01}$ & $\mathbf{94.80 \pm 0.05}$\\
      \bottomrule
    \end{tabular}
    \label{tab:results}
\end{table}

\begin{table}[ht!]
    \centering
    \caption{\small{Accuracy of switching the agents' partners. We also did ablation study for our model by replacing belief over belief in agent Q-functions with a hidden variable output from GRU (ToM-noBoB). ToM with partner's belief modeling has the highest performance in both experiments. BAD's performance drop proves our analysis of weakly centralized training and group-specific protocol. All models including ToM witness significant performance decrease for appointment scheduling. Since each agent only has 256 time tables to choose from, agents' action patterns toward calendars are easier to remember and take advantage of by their partners. So, a portion of the emerged protocol is inevitably group-specific. Yet, ToM still managed to form the most universal protocol.}}
    \begin{tabular}{c|c|c}
        \toprule
         & Kitchen & Appointment \\
        \midrule
        BAD & 37.97 $\pm$ 1.76 & 46.89 $\pm$ 3.16 \\
        CIRL & 49.65 $\pm$ 2.16 & N/A\\
        COMA & 43.90 $\pm$ 9.98 & 52.32 $\pm$ 2.00 \\
        DDRQN & 27.20 $\pm$ 2.19 & 42.38 $\pm$ 4.62 \\
        MADDPG & 15.18 $\pm$ 2.35 & 31.32 $\pm$ 13.99 \\
        ToM-noBoB & 87.96 $\pm$ 0.64 & 61.18 $\pm$ 1.66\\
        ToM & $\mathbf{92.51 \pm 0.95}$ & $\mathbf{70.29 \pm 4.09}$ \\
        \bottomrule
    \end{tabular}
    \label{tab:switch_group}
    \vspace{-10pt}
\end{table}
\subsection{Appointment Scheduling}\label{sec:calendar}
In the kitchen collaboration game, only the chef has a private agent state. We now propose a game where both agents have private information. Suppose there are two agents $A$ and $B$, each having a private time table, and they want to schedule a meeting time available to both of them. We code private time table as a $D$-dimension binary vector with 0 meaning free and 1 meaning occupied. Both agents can perform three types of actions, inform, propose and reject. Rejecting ends the game indicating no common available time to the agents. Proposing ends the game indicating a time slot to meet. Informing stands for the speaker informs the other agent that an interval is occupied and unable to meet. Notice that we regulate the message space for informing to avoid trivial conversation. That is, agents are only allowed to say continuous occupied intervals. For instance, if agent $A$ has a schedule $[0, 0, 1, 1, 1, 0, 1, 1]$, then it is allowed to say $(2), (3), (4), (2, 3), (3, 4), (2, 3, 4), (6), (7)$, and $(6, 7)$, but not $(2, 4), (4, 6), (0, 2)$ etc. Rejecting meetable schedules, proposing occupied time slots or informing wrong messages\footnote{Agents cannot send invalid information. Here wrong messages mean valid but wrong messages. For example, $(0, 1, 2)$ in the previous example, because slot 0 and 1 are not occupied. Numbers are 0-index in the example.} all lead to game failure and every informing has a message cost to prevent long-lasting games. At the end of each game, both agents get +1 or -2 for success and failure, respectively, and every correct message has a cost of -0.1. Failures caused by all reasons: wrong proposal, rejection or message, trigger the same failure reward. In this experiment, agents' beliefs should be $2^{D}$-dimensional beliefs. Yet, even if $D = 8$, they are high-dimensional vectors, so we use a $D$-dimensional vector as a concatenated belief, with each dimension representing the probability for partner's that slot being occupied\footnote{This is only for $b_{-i}$, we still use $2^D$-dimensional beliefs for $b_i$.}. Continuous vectors are discretized to be supervisions by rounding each number to the closest decimal, eg. $[0.43, 0.67]$ to $[0.4, 0.7]$. The distance function used for $L^f$ is L2-norm between the supervision vector and the predicted concatenated belief.

Both agents' private time tables are randomly generated with each slot sampled from a Bernoulli distribution with $p = 0.5$. As there are only $2^{2D}$ games, to prevent overfitting, we make sure all schedules in the testing sets are not included in the training sets. There are about 180 schedules, 30k games, and 80 schedules, 6k games in the training and testing sets respectively. We used $D = 8$ in our experiments. Notice that the baseline of random guess accuracy is 25\%; the baseline of random propose based on self time table is 50\%. Our results and analysis are reported in captions of table \ref{tab:results} and figure \ref{fig:calendar8}.

\subsection{Flexible Group Assignment}\label{sec:switch}

Another contribution of our work is that with ToM considered in action selections, universal protocols compatible cross groups are more likely to be developed. That is to say, our model can integrate task level essence into the protocol so that even if the partner in testing is different from that in training, they can still achieve good performance using their protocols learned separately. Task level universal protocol is the prerequisite of flexible agents assignment. Suppose there is a group of agents collaborating to complete tasks. If one of them is broken and needs to be substituted, we expect another agent to take the broken agent's place at once. The new agent doesn't need to have experience working with this particular group if they have a task level protocol. Most centralized training approaches only learn group-specific protocols. 

Since for each split, we run two experiments, we switch the agents between the two experiments. Let agent $A_1$ trained with agent $B_1$ to collaborate with agent $B_2$ and let agent $A_2$ to collaborate with agent $B_1$ in testing. Only when agents understand the task rather than form group-specific tacit agreements can they maintain the performance after switch partners. See table \ref{tab:switch_group} for results. Methods with centralized structures like BAD and COMA suffer from significant performance deduction, revealing the limitation of their group protocols.

\section{Conclusion}
In this paper, we proposed an adaptive multiagent learning algorithm, with which cooperative agents can develop effective collaborations in MAS with imperfect information. Agents learn how to infer others' hidden mental states with only observations of partners' actions and no verbal communications. Our algorithm outperforms all other decentralized execution approaches and shows the least performance drop in group switching experiments, demonstrating that our agents form strategies for the game instead of emerging ad-hoc protocols only compatible to specific partners. In the future, we aim to explore games involving continuous agent state space, where beliefs cannot be modeled as vectors but have to be parameterized.

\bibliographystyle{IEEEtran}
\bibliography{reference}

\begin{thebibliography}{10}
\providecommand{\url}[1]{#1}
\csname url@samestyle\endcsname
\providecommand{\newblock}{\relax}
\providecommand{\bibinfo}[2]{#2}
\providecommand{\BIBentrySTDinterwordspacing}{\spaceskip=0pt\relax}
\providecommand{\BIBentryALTinterwordstretchfactor}{4}
\providecommand{\BIBentryALTinterwordspacing}{\spaceskip=\fontdimen2\font plus
\BIBentryALTinterwordstretchfactor\fontdimen3\font minus
  \fontdimen4\font\relax}
\providecommand{\BIBforeignlanguage}[2]{{%
\expandafter\ifx\csname l@#1\endcsname\relax
\typeout{** WARNING: IEEEtran.bst: No hyphenation pattern has been}%
\typeout{** loaded for the language `#1'. Using the pattern for}%
\typeout{** the default language instead.}%
\else
\language=\csname l@#1\endcsname
\fi
#2}}
\providecommand{\BIBdecl}{\relax}
\BIBdecl

\bibitem{foerster2016learning}
J.~Foerster, I.~A. Assael, N.~de~Freitas, and S.~Whiteson, ``Learning to
  communicate with deep multi-agent reinforcement learning,'' in \emph{Advances
  in Neural Information Processing Systems}, 2016, pp. 2137--2145.

\bibitem{sukhbaatar2016learning}
S.~Sukhbaatar, R.~Fergus \emph{et~al.}, ``Learning multiagent communication
  with backpropagation,'' in \emph{Advances in Neural Information Processing
  Systems}, 2016, pp. 2244--2252.

\bibitem{foerster2017stabilising}
J.~Foerster, N.~Nardelli, G.~Farquhar, T.~Afouras, P.~H. Torr, P.~Kohli, and
  S.~Whiteson, ``Stabilising experience replay for deep multi-agent
  reinforcement learning,'' in \emph{Proceedings of the 34th International
  Conference on Machine Learning-Volume 70}.\hskip 1em plus 0.5em minus
  0.4em\relax JMLR. org, 2017, pp. 1146--1155.

\bibitem{lowe2017multi}
R.~Lowe, Y.~Wu, A.~Tamar, J.~Harb, O.~P. Abbeel, and I.~Mordatch, ``Multi-agent
  actor-critic for mixed cooperative-competitive environments,'' in
  \emph{Advances in Neural Information Processing Systems}, 2017, pp.
  6379--6390.

\bibitem{foerster2016riddle}
J.~N. Foerster, Y.~M. Assael, N.~de~Freitas, and S.~Whiteson, ``Learning to
  communicate to solve riddles with deep distributed recurrent q-networks,''
  \emph{arXiv preprint arXiv:1602.02672}, 2016.

\bibitem{foerster2018bayesian}
J.~N. Foerster, F.~Song, E.~Hughes, N.~Burch, I.~Dunning, S.~Whiteson,
  M.~Botvinick, and M.~Bowling, ``Bayesian action decoder for deep multi-agent
  reinforcement learning,'' \emph{arXiv preprint arXiv:1811.01458}, 2018.

\bibitem{foerster2018counterfactual}
J.~N. Foerster, G.~Farquhar, T.~Afouras, N.~Nardelli, and S.~Whiteson,
  ``Counterfactual multi-agent policy gradients,'' in \emph{Thirty-Second AAAI
  Conference on Artificial Intelligence}, 2018.

\bibitem{premack1978does}
D.~Premack and G.~Woodruff, ``Does the chimpanzee have a theory of mind?''
  \emph{Behavioral and brain sciences}, vol.~1, no.~4, pp. 515--526, 1978.

\bibitem{yoshida2008game}
W.~Yoshida, R.~J. Dolan, and K.~J. Friston, ``Game theory of mind,'' \emph{PLoS
  computational biology}, vol.~4, no.~12, p. e1000254, 2008.

\bibitem{baker2017rational}
C.~L. Baker, J.~Jara-Ettinger, R.~Saxe, and J.~B. Tenenbaum, ``Rational
  quantitative attribution of beliefs, desires and percepts in human
  mentalizing,'' \emph{Nature Human Behaviour}, vol.~1, no.~4, p. 0064, 2017.

\bibitem{de2015higher}
H.~De~Weerd, R.~Verbrugge, and B.~Verheij, ``Higher-order theory of mind in the
  tacit communication game,'' \emph{Biologically Inspired Cognitive
  Architectures}, vol.~11, pp. 10--21, 2015.

\bibitem{kinney1998learning}
M.~Kinney and C.~Tsatsoulis, ``Learning communication strategies in multiagent
  systems,'' \emph{Applied intelligence}, vol.~9, no.~1, pp. 71--91, 1998.

\bibitem{das2017learning}
A.~Das, S.~Kottur, J.~M. Moura, S.~Lee, and D.~Batra, ``Learning cooperative
  visual dialog agents with deep reinforcement learning,'' in \emph{Proceedings
  of the IEEE International Conference on Computer Vision}, 2017, pp.
  2951--2960.

\bibitem{doshi2009monte}
P.~Doshi and P.~J. Gmytrasiewicz, ``Monte carlo sampling methods for
  approximating interactive pomdps,'' \emph{Journal of Artificial Intelligence
  Research}, vol.~34, pp. 297--337, 2009.

\bibitem{han2018learning}
Y.~Han and P.~Gmytrasiewicz, ``Learning others' intentional models in
  multi-agent settings using interactive pomdps,'' in \emph{Advances in Neural
  Information Processing Systems}, 2018, pp. 5634--5642.

\bibitem{de2014theory}
H.~De~Weerd, R.~Verbrugge, and B.~Verheij, ``Theory of mind in the mod game: An
  agent-based model of strategic reasoning.'' in \emph{ECSI}, 2014, pp.
  128--136.

\bibitem{de2017estimating}
H.~De~Weerd, D.~Diepgrond, and R.~Verbrugge, ``Estimating the use of
  higher-order theory of mind using computational agents,'' \emph{The BE
  Journal of Theoretical Economics}, vol.~18, no.~2, 2017.

\bibitem{claus1998dynamics}
C.~Claus and C.~Boutilier, ``The dynamics of reinforcement learning in
  cooperative multiagent systems,'' \emph{AAAI/IAAI}, vol. 1998, pp. 746--752,
  1998.

\bibitem{fisac2017pragmatic}
J.~F. Fisac, M.~A. Gates, J.~B. Hamrick, C.~Liu, D.~Hadfield-Menell,
  M.~Palaniappan, D.~Malik, S.~S. Sastry, T.~L. Griffiths, and A.~D. Dragan,
  ``Pragmatic-pedagogic value alignment,'' \emph{arXiv preprint
  arXiv:1707.06354}, 2017.

\bibitem{williams1992simple}
R.~J. Williams, ``Simple statistical gradient-following algorithms for
  connectionist reinforcement learning,'' \emph{Machine learning}, vol.~8, no.
  3-4, pp. 229--256, 1992.

\bibitem{watkins1989learning}
C.~J. C.~H. Watkins, ``Learning from delayed rewards,'' Ph.D. dissertation,
  King's College, Cambridge, 1989.

\bibitem{hernandez2017survey}
P.~Hernandez-Leal, M.~Kaisers, T.~Baarslag, and E.~M. de~Cote, ``A survey of
  learning in multiagent environments: Dealing with non-stationarity,''
  \emph{arXiv preprint arXiv:1707.09183}, 2017.

\bibitem{yang2004multiagent}
E.~Yang and D.~Gu, ``Multiagent reinforcement learning for multi-robot systems:
  A survey,'' tech. rep, Tech. Rep., 2004.

\bibitem{bu2008comprehensive}
L.~Bu, R.~Babu, B.~De~Schutter \emph{et~al.}, ``A comprehensive survey of
  multiagent reinforcement learning,'' \emph{IEEE Transactions on Systems, Man,
  and Cybernetics, Part C (Applications and Reviews)}, vol.~38, no.~2, pp.
  156--172, 2008.

\bibitem{tampuu2017multiagent}
A.~Tampuu, T.~Matiisen, D.~Kodelja, I.~Kuzovkin, K.~Korjus, J.~Aru, J.~Aru, and
  R.~Vicente, ``Multiagent cooperation and competition with deep reinforcement
  learning,'' \emph{PloS one}, vol.~12, no.~4, p. e0172395, 2017.

\bibitem{mordatch2018emergence}
I.~Mordatch and P.~Abbeel, ``Emergence of grounded compositional language in
  multi-agent populations,'' in \emph{Thirty-Second AAAI Conference on
  Artificial Intelligence}, 2018.

\bibitem{lazaridou2016multi}
\BIBentryALTinterwordspacing
A.~Lazaridou, A.~Peysakhovich, and M.~Baroni, ``Multi-agent cooperation and the
  emergence of (natural) language,'' in \emph{International Conference on
  Learning Representations}, 2017. [Online]. Available:
  \url{https://openreview.net/forum?id=Hk8N3Sclg}
\BIBentrySTDinterwordspacing

\bibitem{havrylov2017emergence}
S.~Havrylov and I.~Titov, ``Emergence of language with multi-agent games:
  Learning to communicate with sequences of symbols,'' in \emph{Advances in
  neural information processing systems}, 2017, pp. 2149--2159.

\bibitem{evtimova2017emergent}
K.~Evtimova, A.~Drozdov, D.~Kiela, and K.~Cho, ``Emergent language in a
  multi-modal, multi-step referential game,'' \emph{arXiv preprint
  arXiv:1705.10369}, 2017.

\bibitem{lazaridou2018emergence}
\BIBentryALTinterwordspacing
A.~Lazaridou, K.~M. Hermann, K.~Tuyls, and S.~Clark, ``Emergence of linguistic
  communication from referential games with symbolic and pixel input,'' in
  \emph{International Conference on Learning Representations}, 2018. [Online].
  Available: \url{https://openreview.net/forum?id=HJGv1Z-AW}
\BIBentrySTDinterwordspacing

\bibitem{wagner2003progress}
K.~Wagner, J.~A. Reggia, J.~Uriagereka, and G.~S. Wilkinson, ``Progress in the
  simulation of emergent communication and language,'' \emph{Adaptive
  Behavior}, vol.~11, no.~1, pp. 37--69, 2003.

\bibitem{foerster2018learning}
J.~Foerster, R.~Y. Chen, M.~Al-Shedivat, S.~Whiteson, P.~Abbeel, and
  I.~Mordatch, ``Learning with opponent-learning awareness,'' in
  \emph{Proceedings of the 17th International Conference on Autonomous Agents
  and MultiAgent Systems}.\hskip 1em plus 0.5em minus 0.4em\relax International
  Foundation for Autonomous Agents and Multiagent Systems, 2018, pp. 122--130.

\bibitem{wen2019probabilistic}
Y.~Wen, Y.~Yang, R.~Luo, J.~Wang, and W.~Pan, ``Probabilistic recursive
  reasoning for multi-agent reinforcement learning,'' \emph{arXiv preprint
  arXiv:1901.09207}, 2019.

\bibitem{wei2018multiagent}
E.~Wei, D.~Wicke, D.~Freelan, and S.~Luke, ``Multiagent soft q-learning,'' in
  \emph{2018 AAAI Spring Symposium Series}, 2018.

\bibitem{haarnoja2017reinforcement}
T.~Haarnoja, H.~Tang, P.~Abbeel, and S.~Levine, ``Reinforcement learning with
  deep energy-based policies,'' in \emph{Proceedings of the 34th International
  Conference on Machine Learning-Volume 70}.\hskip 1em plus 0.5em minus
  0.4em\relax JMLR. org, 2017, pp. 1352--1361.

\bibitem{wang2016learning}
D.~Wang and Q.~Liu, ``Learning to draw samples: With application to amortized
  mle for generative adversarial learning,'' \emph{arXiv preprint
  arXiv:1611.01722}, 2016.

\bibitem{jaakkola1994convergence}
T.~Jaakkola, M.~I. Jordan, and S.~P. Singh, ``Convergence of stochastic
  iterative dynamic programming algorithms,'' in \emph{Advances in neural
  information processing systems}, 1994, pp. 703--710.

\bibitem{choi2018compositional}
E.~Choi, A.~Lazaridou, and N.~de~Freitas, ``Compositional obverter
  communication learning from raw visual input,'' \emph{arXiv preprint
  arXiv:1804.02341}, 2018.

\bibitem{mnih2013playing}
V.~Mnih, K.~Kavukcuoglu, D.~Silver, A.~Graves, I.~Antonoglou, D.~Wierstra, and
  M.~Riedmiller, ``Playing atari with deep reinforcement learning,''
  \emph{arXiv preprint arXiv:1312.5602}, 2013.

\bibitem{hausknecht2015deep}
M.~Hausknecht and P.~Stone, ``Deep recurrent q-learning for partially
  observable mdps,'' in \emph{2015 AAAI Fall Symposium Series}, 2015.

\end{thebibliography}
\end{document}